\documentclass[aps,amsfonts,pra,twocolumn,showpacs,longbibliography]{revtex4-1}

\usepackage{subfigure}
\usepackage{textcomp}
\usepackage{graphicx}
\usepackage{amssymb}
\usepackage{amsmath}
\usepackage{bm}
\usepackage{amsmath, amsthm, amssymb}
\usepackage{color}
\usepackage[colorlinks=true,linkcolor=blue,urlcolor=blue,citecolor=blue]{hyperref}

\newcommand{\eq}[1]{Eq.~(\ref{#1})}

\newcommand{\fig}[1]{Fig.\thinspace{}\ref{#1}}
\newcommand{\figg}[2]{Fig.\thinspace{}\ref{#1} and \ref{#2}}
\newcommand{\fc}[1]{({#1})}
\newcommand{\figc}[2]{Fig.\thinspace{}\ref{#1}\thinspace{}\fc{#2}}

\newcommand{\mkch}[1]{{\color{black} #1}}

\def\bra#1{\mathinner{\langle{#1}|}}
\def\ket#1{\mathinner{|{#1}\rangle}}

\def\beq{\begin{equation}}
\def\eeq{\end{equation}}
\def\bea{\begin{eqnarray}}
\def\eea{\end{eqnarray}}

\begin{document}

\title{Regimes of heating and dynamical response in driven many-body localized systems}

\author{Sarang Gopalakrishnan}
\affiliation{Department of Physics and Walter Burke Institute, California Institute of Technology, Pasadena, CA 91125, USA}%

\author{Michael Knap}
\affiliation{Department of Physics, Walter Schottky Institute, and Institute for Advanced Study, Technical University of Munich, 85748 Garching, Germany}%

\author{Eugene Demler}
\affiliation{Department of Physics, Harvard University, Cambridge, MA 02138, USA}%

\begin{abstract}
We explore the response of many-body localized (MBL) systems to periodic driving of arbitrary amplitude, focusing on the rate at which they exchange energy with the drive. To this end, we introduce an infinite-temperature generalization of the effective ``heating rate'' in terms of the spread of a random walk in energy space. We compute this heating rate numerically and estimate it analytically in various regimes. When the drive amplitude is much smaller than the frequency, this effective heating rate is given by linear response theory with a coefficient that is proportional to the optical conductivity; in the opposite limit, the response is nonlinear and the heating rate is a nontrivial power-law of time. We discuss the mechanisms underlying this crossover in the MBL phase. We comment on implications for the subdiffusive thermal phase near the MBL transition, and for response in imperfectly isolated MBL systems. 
\end{abstract}

\maketitle

\section{Introduction}

Isolated quantum systems in the many-body localized (MBL) phase do not approach local thermal equilibrium starting from generic initial conditions~\cite{pwa, BAA, Gornyi, nhreview, altman_universal_2015}. Instead, in the MBL phase, transport and relaxation are absent, and a system retains memory of its initial conditions at arbitrarily late times. At present, there is strong evidence---from numerical studies~\cite{oh, znidaric2008, monthus2010, PalHuse, berkreich}, rigorous mathematical approaches~\cite{jzi}, and experiments~\cite{ovadia2014evidence, demarco2015, schreiber2015observation, smith2015, bordia2015}---that the MBL phase exists in strongly disordered one-dimensional spin and fermion systems. Moreover, a phenomenological description exists for systems deep in the MBL phase~\cite{spa_LIOM, hno, ros2015integrals, rademaker2016, pollmann2015efficient, khemani2015obtaining, pekker2015}, and can be used to explore aspects of dynamics and response~\cite{bpm, spa_entanglement, ngh, DEER, serbyn_quench, vpm, mbmott, xu2015response}. Recently, the \emph{transition} between MBL and thermal phases has also been explored, using general arguments~\cite{grover2014,blcr, Agarwal, clo, gopalakrishnan_griffiths_2015}, mean-field theory~\cite{gn}, and renormalization-group schemes~\cite{VHA, PVPtransition, trithep}. The nature of this transition, and the MBL phase, is of particular interest because equilibrium statistical mechanics fails at the transition and does not apply in the MBL phase. Thus, we might expect various features of dynamics and response in the MBL phase to differ dramatically from equilibrium expectations.

\begin{figure}[b]
\begin{center}
\includegraphics[width=.48\textwidth]{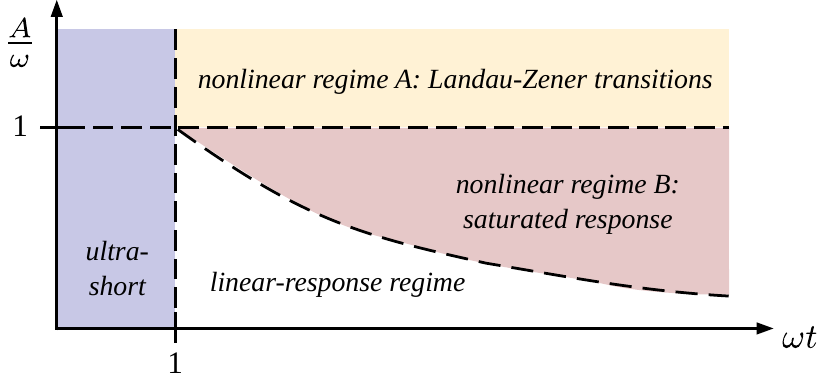}
\caption{Heating regimes and dynamical response to periodic driving in the many-body localized phase as a function of driving strength $A$ and time $t$, at fixed driving frequency $\omega$. At time $t < 1/\omega$, the response is in the ``ultrashort-time'' regime: the driving frequency cannot be resolved and the heating rate of the system is protocol-dependent. At times such that $1/\omega < t < 1/A$, resonant transitions govern heating and the rate is given by linear response. At times $t > 1/A$, the resonant transitions are saturated but slower processes (Sec.~\ref{se:intdrive}) still contribute to heating. When $A>\omega$, the linear-response window vanishes and heating is given by Landau-Zener transitions. We find numerically and argue analytically that response in this regime is nonlinear in time; we furthermore predict that its amplitude-dependence is inconsistent with linear response theory.}
\label{pdmain}
\end{center}
\end{figure}

The present work addresses one such exotic feature of MBL systems: namely, that in these systems, the d.c. limit of response functions is ill-defined. For concreteness, consider the conductivity of the system, i.e., its response to periodic driving by an electric field of amplitude $A$ and frequency $\omega$. In a typical thermalizing phase, this response is linear in $A$, for small enough $A$, regardless of the drive frequency: the linear response limit $A \rightarrow 0$ and the d.c. limit $\omega \rightarrow 0$ commute. However, in the MBL phase, these limits do \emph{not} commute~\cite{khemani2014}. Taking the limit $A \rightarrow 0$ at fixed finite frequency gives rise to the linear-response optical conductivity $\sigma(\omega) \sim \omega^\alpha$ discussed in Ref.~\cite{mbmott}, which vanishes as $\omega \rightarrow 0$. On the other hand, taking the $\omega \rightarrow 0$ limit at fixed $A$ gives rise to a drive-induced many-body delocalization transition~\cite{ldm, abanin2014theory}, and therefore a breakdown of linear response theory~\cite{khemani2014}. 

The objective of this work is to study response in the MBL phase beyond these two limits, for general $A/\omega$ (but provided these are small compared with the intrinsic energy scales of the system; the opposite case is addressed in Ref.~\cite{Kozarzewski,Pollmann}). Our main results are as follows. We identify an observable---specifically, a generalized heating rate---that can be numerically extracted from the dynamics of the driven isolated system. This heating rate allows us to characterize dynamical response without relying on linear response theory (which breaks down as $\omega \rightarrow 0$). We then identify the processes that dominate heating and response for various regimes of $A/\omega$, arguing that linear response is due to absorption from resonant configuration-pairs~\cite{mbmott} and occurs in a time-window $1/\omega \alt t \alt 1/A$ (starting from when the drive is turned on). These processes give rise to the expected Joule-heating behavior, in which the energy absorbed (or, equivalently, the dissipated power) $\sim A^2 \sigma(\omega)$. Linear-response processes saturate on timescales $t \agt 1/A$, but subleading processes still contribute slow dynamics. For stronger drive, we identify Landau-Zener transitions (and, potentially, thermal Griffiths inclusions) as the dominant contributor to response. These mechanisms cause heating that is a nontrivial power-law of both time and drive amplitude. The associated exponents vary continuously through the MBL phase. The various regimes are sketched in Fig.~\ref{pdmain}. We support our heuristic analytical estimates with numerical evidence.

This work is organized as follows. First, we introduce a scheme for computing the heating rate in Sec.~\ref{se:dynresp}. Then we review the effective spin model describing the MBL phase in Sec~\ref{model}, and discuss the various regimes of heating and their relevant scales in Sec.~\ref{se:regimes}. The transient linear-response regime is studied analytically in Sec.~\ref{lr} and the non-linear dynamics in Sec.~\ref{se:intdrive}. In Sec.~\ref{se:num} we numerically demonstrate the various regimes of linear and non-linear response. Finally, in Sec.~\ref{se:dis}, we comment on experimental implications and possible extensions of our analysis, particularly to the case of imperfectly isolated systems.

\section{Measurement of dynamical response}\label{se:dynresp}

In this work, we are interested in understanding the regime of validity of linear response theory and the crossover to non-linear response. Therefore, we cannot compute the conductivity using the Kubo formula (as in Ref.~\cite{mbmott}) because this \emph{relies} on linear response: our objective, in other words, is to \emph{determine} where the Kubo formula works and how the system responds beyond that regime. Thus, we need to compute response \emph{directly} from the behavior of a driven system.

In general, one determines the conductivity of a system by applying a perturbatively small time-dependent electric field and measuring the response of the current to that perturbation. Implicitly, this standard definition assumes that the perturbed system has reached a steady state, e.g., because it is coupled to a heat bath that dissipates energy. Applying this standard notion to the MBL context raises the following difficulty: we are interested in systems that are isolated from the environment on the timescales of interest, so there is no external source of dissipation to bring the system to its steady state. One must instead extract the conductivity from a \emph{transient}: specifically, one can extract the conductivity from the dissipated power, or Joule-heating rate (given by $V^2 G$, where $G$ is the conductance and $V$ the applied voltage), when a system is driven starting at some time $t = 0$. A practical challenge with computing heating rates, however, is that the regime of interest to us is one of high or even infinite temperature of the system. In this infinite-temperature limit, the amount of heating is necessarily small, which makes direct numerical extraction of heating rates challenging~\footnote{Heating from the ground state is considered in Refs.~\cite{Kozarzewski, Pollmann}. Note that the linear-response conductivity at zero temperature is the same for MBL and noninteracting systems~\cite{mbmott}, and that the discussion of Ref.~\cite{khemani2014} is also greatly modified in this limit. Thus, the results of Refs.~\cite{Kozarzewski, Pollmann} do not directly address the conceptual issues that are relevant to the present work.}.

One can address this difficulty by thinking about the mechanics of the heating process. Suppose the system is initially in an eigenstate in the middle of the many-body spectrum. During a particular drive cycle, the system is equally likely to absorb or to emit a quantum of the drive. Thus, the energy of the system undergoes a random walk, with a step set by the drive frequency $\omega$. When the system is instead initialized \emph{near} infinite temperature (i.e., at a temperature $T$ greater than the intrinsic system scales and drive frequency), then the initial occupation of an eigenstate (in the eigenbasis of the undriven Hamiltonian) is given by $\approx 1 - E/T$, where $E$ is the energy. As states with lower energy are slightly more likely to be initially occupied, \emph{on average} the random walk causes energy to be gained and the system heats up; i.e., the energy space initially has a ``concentration gradient'' (proportional to $1/T$) and ``heating'' results from the dynamics relaxing this initial gradient (see App.~\ref{gradients}).
Thus, it is plausible that, up to a factor of $T$, the heating rate in the high-temperature limit is related to the fictitious \emph{diffusion constant} in energy space. Indeed, this connection is well-understood for the thermal phase~\cite{jarzynski1997, cohen1999}. 

This fictitious diffusion constant has a nonzero limit at infinite temperature, and is easy to measure numerically, by initializing the system in an eigenstate (or a wavepacket with narrow energy spread) and measuring the energy spread of the wavepacket as a function of time. Specifically, we introduce the energy spread $(\Delta E)^2$, as 

\bea
(\Delta E)^2  \equiv  \langle m(t) | \hat{H}^2 | m(t) \rangle  - \langle m(t) |\hat{H}  | m(t) \rangle^2. \label{eq:enspread}
\eea
Here,  $\ket{m(t)}=\hat U(t) \ket{m}$ is the time evolved state, with $\ket{m}$ being an eigenstate of the unperturbed Hamiltonian $\hat H$ and $\hat U(t)$ the unitary time evolution operator generated by $\hat H + \hat H_\text{drv.}(t)$. We emphasize that this ``fictitious'' diffusion constant is distinct from, and not directly related to, the ``true'' energy diffusion constant of the undriven system: the ``fictitious'' energy diffusion constant measures the spread of \emph{probability} in \emph{Fock} space, whereas the ``true'' energy diffusion constant measures the spread of \emph{energy} in \emph{real} space. For a driven system, energy is not conserved (and thus the true energy diffusion constant is not physically meaningful) whereas probability is conserved, so the fictitious diffusion constant remains meaningful. (We note that the fictitious diffusion constant is also conceptually related to the quantum Fisher information~\cite{smith2015, caves1994}.)

In what follows, we shall be primarily interested in computing the dynamics of the observable $(\Delta E)^2$ in Eq.~\eqref{eq:enspread}. We believe that for local systems this quantity is quite generally proportional to the high-temperature limit of the heating rate. This proportionality is known to exist in classical chaotic systems~\cite{jarzynski1997} as well as their quantum equivalents~\cite{cohen1999}.
We also show \emph{explicitly} that the relation holds for an MBL system driven very weakly at a nonzero frequency (i.e., $A/\omega \ll 1$): 
$(\Delta E)^2$ grows linearly with time $t$, with a coefficient that is $\sim T \sigma(\omega) A^2$, where $\sigma(\omega)$ is the linear-response a.c. conductivity~\cite{mbmott}, i.e., $(\Delta E)^2 \sim T \sigma(\omega) A^2 t$. This corresponds precisely to the Joule heating rate of a system with conductivity $\sigma(\omega)$. The correspondence between heating rates and energy spread can also be shown generally for systems in which heating is due to isolated two-level systems (App.~\ref{gradients}). 

The linear response regime can fail either through violations of the $A^2$ dependence or because the $t$-dependence ceases to be linear, for example because of saturation effects. We shall discuss these effects in more detail below but first we introduce the effective-spin model describing the many-body localized phase.

\section{Effective-spin model in the many-body localized phase}\label{model}

We consider one-dimensional systems, described by local Hamiltonians (e.g., the random-field Heisenberg chain, see Eq.~\eqref{eq:h} below) and focus on the regime where all many-body eigenstates are in the MBL phase. In this regime, a phenomenological description of the system exists, in terms of effective spins-1/2 labeled $\tau^z_i$ (also known as local integrals of motion or l-bits~\cite{spa_LIOM, hno, ros2015integrals, rademaker2016}):

\beq\label{lbit}
\hat H = \sum_i h_i \tau^z_i + \sum_{ij} J_{ij} \tau^z_i \tau^z_j + \sum_{ijk} K_{ijk} \tau^z_i \tau^z_j \tau^z_k + \ldots
\eeq
The effective degrees of freedom $\tau^z_i$ are related to the microscopic ones (denoted $\hat S_i^\alpha$) by a finite-depth unitary transformation~\cite{BauerNayak}, up to exponential tails. For notational simplicity (and to make contact with numerics) we shall work in one dimension, with open boundary conditions; none of our considerations relies crucially on these assumptions. Then the time-varying electric field can be written as $\hat H_\text{drv.} =A\sin \omega t \sum\nolimits_i x_i \hat S^z_i$.

The expansion of a particular $\hat S$ operator, e.g., $\hat S^x$, in terms of $\tau$ operators, has the form $\hat S^x_i \simeq \sum\nolimits F^{1\alpha}_{ij} \tau^\alpha_j + F^{2, \alpha\beta}_{ijk} \tau^\alpha_j \tau^\beta_k + \ldots$. The $F$ coefficients fall off exponentially with the furthest distance between the $\tau$ spins involved, and also fall off exponentially with the number of \emph{off-diagonal} $\tau$ operators (i.e., $\tau^x$ or $\tau^y$) involved~\cite{gn, mbmott}. For example, the coefficient of a term of the form $\prod\nolimits_{p = 1}^{m} \tau^{x, y}_{i_p} \prod\nolimits_{q = m + 1}^{n} \tau^z_{i_q}$ would fall off as $\exp(-x/\xi - m/\zeta)$, where $x \equiv \mathrm{max}(|i_p - i_{p'}|)$. Stability of the MBL phase at infinite temperature requires that $s\zeta  < 1$~\cite{BAA, gn, mbmott} as the available phase space for $m$ spin-flips grows as $\exp [{s m}]$. At infinite temperature the entropic factor $s\sim \log 2$.

\section{Regimes of heating and relevant scales}\label{se:regimes}

In this section, we qualitatively introduce the two primary heating mechanisms: resonant transitions and Landau-Zener transitions. We then identify regimes in which each mechanism is dominant, and explore the implications for heating in those regimes.

\subsection{Resonant transitions}\label{restrans}

We first consider what happens when one drives the Hamiltonian~\eqref{lbit} very weakly at relatively high frequency, $A/\omega\ll 1$. We assume that the drive is turned on instantaneously at time $t = 0$, and that the system is initialized in a many-body eigenstate, i.e., in a product state of the effective spins $\tau_i$. The drive is diagonal in the \emph{physical} spin basis; thus, in the \emph{effective} spin basis, the drive generically has off-diagonal matrix elements for rearranging multiple effective spins. These typically fall off exponentially with order and inter-spin distance (as discussed in the previous section). However, there are rare pairs of effective-spin configurations between which the drive has a large matrix element. For an illustrative example, consider a well-localized Anderson insulator. Most (single-particle) eigenstates in the Anderson insulator are localized on single sites; however, rare eigenstates are delocalized across a \emph{resonant pair} of accidentally degenerate sites~\cite{mott1968}. The eigenstates in this resonant pair of sites consist of symmetric and antisymmetric combinations of the single-site orbitals, i.e., $|\psi_\pm \rangle = |a \rangle \pm | b \rangle$ where $a$ and $b$ are the two orbitals. The electric field (which in this basis is $\sim |a \rangle \langle a| - |b \rangle \langle b|$) has matrix elements between $|\psi_+\rangle$ and $|\psi_-\rangle$ that \emph{grow} with the distance between the two sites. Such resonant pairs exist at all scales, and dominate the linear-response conductivity in both single-particle~\cite{mott1968} and MBL systems~\cite{burin1998, yao2013, mbmott}. However, the number of resonances at scale $x$ falls off exponentially with $x$ whenever the MBL phase is stable. In the MBL phase, such resonant pairs exist not just between different sites, but also between different pairs of \emph{configurations}~\cite{burin1998, yao2013, mbmott}; thus the number of resonant pairs is parametrically larger, but their qualitative physics is not greatly modified. 

In the initial eigenstate of the undriven system, each of these resonant pairs is in either its symmetric or antisymmetric state. When the drive is turned on, it induces transitions between these two states provided the transition is resonant with the drive frequency $\omega$; the associated Rabi frequency is set by $A x_\omega$, where $x_\omega$ is the size of the resonant pair (i.e., the dipole moment of the transition). Thus, on a timescale set by the drive amplitude $A$, these resonant pairs are saturated (i.e., each resonant pair is precessing), and beyond this point there is very little absorption. Note that the saturation of resonant pairs is analogous to the phenomenon of spectral hole-burning in glasses~\cite{black-halperin}. To make contact with the effective-spin language of Sec.~\ref{model}, the occupations of the symmetric and antisymmetric orbitals count as conserved quantities, $\tau^z_\pm$. The drive mixes the two orbitals, and therefore has an off-diagonal matrix element of the form $(\tau^+_+ \tau^-_- + \mathrm{h.c.})$.

We now briefly review the counting~\cite{mbmott} of these resonant pairs in the MBL phase, when the system is driven at frequency $\omega$. For brevity we shall quote and use the result of Ref.~\cite{mbmott} that (at high temperature) the most common resonances at low frequencies involve flipping a substantial fraction ($\sim 1/2$ at infinite temperature) of the effective spins within a region of length $x$. Thus, we shall take $n \sim x$ in what follows.
Now, resonances that flip $n$ effective spins have matrix elements $M \sim W \exp(-n/\zeta)$. Thus, such resonances also have \emph{splittings} (owing to hybridization) $\delta \sim W \exp(-n/\zeta)$, and do not contribute at lower frequencies. By contrast, when two configurations are separated by an energy $\omega$, but the matrix element is $W \exp(-n/\zeta) \ll \omega$, then these configurations will not be resonant. Consequently, resonant transitions at frequency $\omega$ are predominantly those for which $W \exp(-n/\zeta) \simeq \omega$. Taking $n \sim x$, this sets a length-scale for resonant transitions

\beq
x_{\text{Mott}} \sim \zeta \log(W/\omega) \label{xmott}.
\eeq
One might intuitively expect linear-response theory to hold when resonant transitions are dominant, because the transition rate is proportional to $A^2$ owing to the Golden Rule. We shall see below that this is indeed the case.

\subsection{Landau-Zener crossings}\label{introlz}

In addition to resonant pairs, a second class of processes that contribute to heating are Landau-Zener transitions~\cite{shevchenko2010}, which we now discuss. Suppose the system begins in a many-body eigenstate, i.e., a product state, or particular configuration, of the effective $\tau$ spins. The drive has matrix elements that are \emph{diagonal} in the effective-spin basis, and thus change the energies of the various configurations (Sec.~\ref{model}); in addition, it has \emph{off-diagonal} matrix elements that can cause transitions between $\tau$-spin eigenstates. During a typical drive cycle, various configurations cross each other in energy. When such a crossing occurs, there is some probability of an adiabatic transition, i.e., one in which the system switches between configurations (as opposed to a \emph{diabatic} transition, in which the system maintains its initial configuration). The matrix element for an adiabatic transition depends on the real-space and configuration-space distance between the configurations (Sec.~\ref{model}). At longer distances, there are more crossings, but they are less likely to be adiabatic (because the matrix element decreases). Quantitatively, the probability of an adiabatic transition at distance $x$ in the many-body case is given by $P_\mathrm{ad}(x) \sim 1 - \exp[- M^2/(A x \omega)]$, where $M \sim W \exp(-x/\zeta)$ is the matrix element between the configurations
\begin{equation}
 P_\text{ad}(x) \sim 1- \exp[-W^2 e^{-2x/\zeta} /(A x \omega)].
 \label{eq:LZ}
\end{equation}
To get the contribution of these LZ crossings to the heating rate, we must identify the conditions under which they cause heating. During each drive cycle, a given LZ crossing occurs \emph{twice}. If it is crossed adiabatically \emph{or} diabatically on both attempts, the system deterministically returns to its initial configuration at the end of a drive cycle. This does not cause heating. Rather, the rate at which a particular transition causes heating is given by $P_{\text{ad}} (1 - P_{\text{ad}})$~\cite{khemani2014, abanin2014theory}: thus, transitions that cause heating are those that have an appreciable probability of happening diabatically and also an appreciable probability of happening adiabatically~\footnote{Here we have assumed that only two levels are involved, as this is the case of most interest to us. However, one can generalize the principle that dissipation is governed by the probability of the system not returning to its original configuration after a drive cycle.}.

We now estimate $P_{\text{ad}}$ for the crossings that \emph{typically} occur when the system is driven with amplitude $A$. Let us consider a segment of size $x$. An electric field of amplitude $A$ shifts energy levels by an amount $\sim A x$. The number of configurations of the effective spins in this segment is $\exp(sx)$ [specifically, $2^x$ at infinite temperature], and their energy bandwidth is $W x$. Thus, if the initial configuration covers an energy window $A x$, it will typically cross $\exp(sx) A/W$ configurations. Thus, in order for at least one LZ transition to typically occur, one needs to look at segments of size 

\beq
x_{\mathrm{LZ}} \sim  (1/s) \log(W/A).\label{xlz}
\eeq

\begin{figure}[t]
\begin{center}
\includegraphics[width=0.45\textwidth]{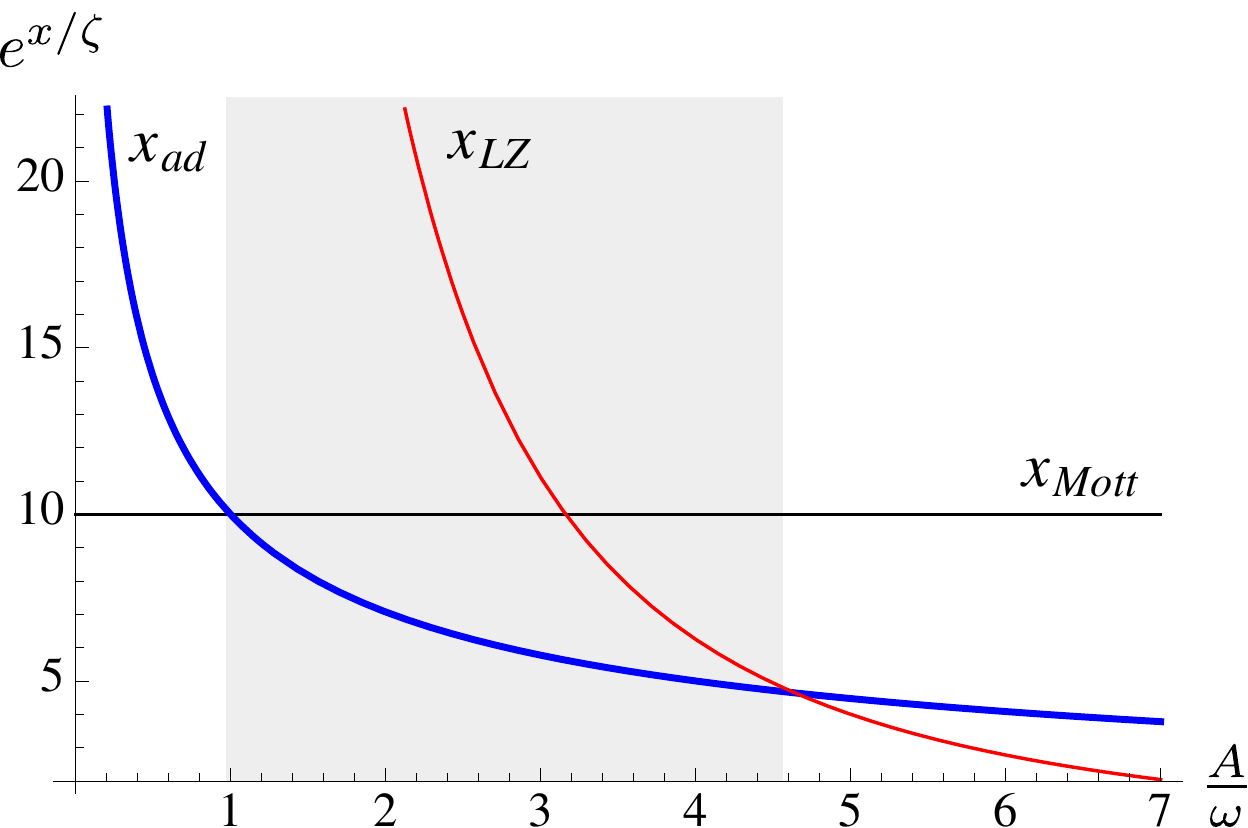}
\caption{Behavior of the length-scales $x_{\text{Mott}}, x_{\text{LZ}}, x_{\text{ad}}$ (defined in text) as the drive amplitude $A$ is varied at fixed frequency $\omega$. We assume $A, \omega \ll W$ (where $W$ is the single-particle bandwidth) and set $s\zeta = 1/2$, so the system is relatively deep in the MBL phase. Thus there are two separate crossovers as $A$ is increased: linear-response fails when $x_{\text{ad}} \simeq x_{\text{Mott}}$, and Landau-Zener transitions percolate when $x_{\text{ad}} \simeq x_{\text{LZ}}$. These crossovers are separated by an intermediate regime (shaded region) in which rare Landau-Zener transitions dominate the response. As $s \zeta$ increases, the intermediate regime shrinks and disappears when $s \zeta = 1$ (i.e., at the MBL transition).}
\label{lengths}
\end{center}
\end{figure}

There are two regimes of behavior depending on whether $P_\mathrm{ad} (x_{\mathrm{LZ}}) \ll 1$ (i.e., most LZ crossings are diabatic) or not. In the limit that $P_\mathrm{ad} (x_{\mathrm{LZ}}) \ll 1/2$, the density of adiabatic LZ transitions per cycle is low. In this case, LZ transitions do not destabilize the MBL phase, but simply provide an additional heating channel in addition to resonant transitions. In the opposite limit, $P_\mathrm{ad} (x_{\mathrm{LZ}}) \sim O(1)$, adiabatic LZ transitions become dense; thus, delocalization takes place through a series of adiabatic LZ hops. This corresponds to a drive-induced many-body delocalization phase transition~\cite{abanin2014theory}. The resulting delocalized phase is presumably thermal (in the sense that it heats up to infinite temperature), but its properties (such as response functions) are not adiabatically connected to those of the undriven system.

\subsection{Length-scales and regimes of response}

\begin{figure}[t]
\begin{center}
\includegraphics[width=.48\textwidth]{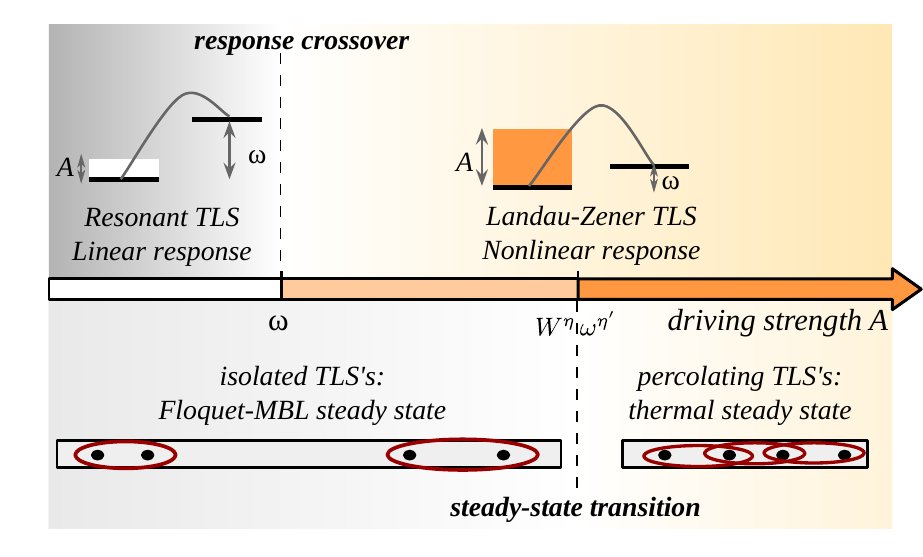}
\caption{Regimes of transient (top) and steady-state behavior (bottom) in driven MBL systems. As the drive strength $A$ is increased at constant frequency $\omega$, there is a crossover between linear and nonlinear response in the transient dynamics, set by the failure of the rotating-wave approximation to the driven two-level systems (TLS's) that govern the response of the system. This crossover happens when these TLS's transition from a drive-resonant regime (top-left) to a Landau-Zener regime (top-right). There is a separate steady-state phase transition between MBL and thermal Floquet Hamiltonians (bottom). This is determined not by the \emph{nature} of TLSs, but rather by the \emph{density} of the dominant type. When TLSs percolate, the steady state is thermal; otherwise it is MBL. The steady-state transition is set by the condition $A^{1 - s\zeta/2} \sim \omega^{s \zeta/2} W^{1 - s \zeta}$. In summary, there are three steady-state regimes for a driven MBL system: (i) MBL long-time behavior with isolated {resonant} transitions; (ii) MBL long-time behavior with isolated {Landau-Zener} transitions; (iii) thermal long-time behavior because of {percolating} Landau-Zener transitions.}
\label{regimefig}
\end{center}
\end{figure}

The previous discussion suggests that there are three separate length-scales governing the response of the system. One of these is the ``Mott'' length-scale, $x_{\text{Mott}} \sim \zeta \log(W/\omega)$, which is the length-scale on which resonant transitions take place [Eq.\eqref{xmott}]. The second is the Landau-Zener crossing scale, $x_{\mathrm{LZ}} \simeq (1/s) \log(W/A)$, which is the distance (in real and/or configuration space) to the nearest Landau-Zener crossing [Eq.~\eqref{xlz}]. Finally, there is a length-scale, which we call the ``adiabatic'' scale, $x_{\mathrm{ad}}$, determined by the condition that $P_{\mathrm{ad}}(x_{\mathrm{ad}}) \sim 1/2$. An approximate formula for this scale is 
\begin{equation}
 x_{\mathrm{ad}} \sim (\zeta/2) \log[W^2/(A \omega)],
\end{equation}
which is obtained by inverting \eq{eq:LZ}. The arguments of Ref.~\cite{abanin2014theory} can be rephrased as saying that when $x_{\mathrm{ad}} \simeq x_{\mathrm{LZ}}$, a drive-induced delocalization transition takes place. The behavior of these three length-scales is shown in Fig.~\ref{lengths}. Note that if $A$ is increased at fixed $\omega$ in the MBL phase, the first crossing that occurs is $x_{\text{ad}} \simeq x_{\text{Mott}}$ when $\omega = A$. At this drive amplitude, $x_{\mathrm{LZ}} > x_{\mathrm{Mott}}, x_{\mathrm{ad}}$ because of the above definitions combined with the condition $s\zeta < 1$, which is required for the stability of the MBL phase, as discussed in Sec.~\ref{model}.

Even before the drive causes delocalization, it causes the breakdown of linear response. The crossover between linear and nonlinear response can be understood as follows (see Fig.~\ref{regimefig}, top): When the drive amplitude is very small $x_{\text{Mott}} \lesssim x_{\mathrm{ad}}$. In that regime, resonant transitions (whose density is set by $\omega$) dominate the response. But as $A$ is ramped up, eventually the phase space for LZ crossings (whose density is set by $A$) dominates that for resonant transitions (even though these LZ crossings have relatively small adiabatic rates). This corresponds to a breakdown of linear response, which is accompanied by a breakdown of the rotating-wave approximation for the driven resonant pairs (cf. top-left and top-right illustrations in Fig.~\ref{regimefig}). In addition to the crossover in the the transient dynamical response, a steady state transition transition from localized to thermal effective Floquet Hamiltonians can be introduced, which is solely set by the density of TLS and not by their character.

There are thus in total three distinct regimes~(\fig{regimefig}): (i) linear response due to isolated resonant TLS's with Floquet steady states that are many-body localized; (ii) nonlinear response due Landau-Zener TLS's, which are nevertheless isolated from one another and hence the steady state remains many-body localized as well (intermediate regime in \figg{lengths}{regimefig}); (iii) nonlinear response due to percolation between TLS's accompanied with thermal steady states induced by strong drive. 

\section{Transient linear response}\label{lr}

This section focuses on regimes in which linear-response behavior emerges. There are two such regimes: in the MBL phase, for sufficiently small $A/\omega$, and in the thermal phase, for general $A/\omega$. We shall consider these in turn. Although our primary concern is with the behavior of the MBL phase, the thermal behavior is instructive and helps to set up our discussion of Griffiths effects in Sec.~\ref{thgriff}.

\subsection{MBL phase: Linear response through resonances}

In this section we analyze a simplified version of the effective-spin model in Sec.~\ref{model}, in which we neglect all degrees of freedom that are not resonant pairs. The two states of each resonant pair can be treated as a two-level system. Note that these resonant two-level systems (RTLS's) are \emph{not} the same as the effective $\tau$ spins in Sec.~\ref{model}, but are much more sparse: most effective $\tau$-spins are not involved in resonances~\footnote{Operationally, a distant or many-spin ``resonant pair'' is a pair of effective spin configurations that are far apart in either real or configuration space, but have a large matrix element of the electric field.}. To emphasize the distinction, we shall denote the RTLS's as $T_\alpha$. Because of their sparseness, we neglect interactions among RTLS's.

We now discuss the dynamics of this ensemble of noninteracting RTLS's. 
We work in the effective spin representation of the undriven system; in the associated natural eigenbasis, each TLS points along $z$ in the absence of drive. The full Hamiltonian of the driven RTLS $\alpha$ can be written as 

\beq
H_\text{RTLS} (\alpha) = \varepsilon_\alpha T^z_\alpha +
2 A \zeta \log(W/\varepsilon_\alpha) \cos(2\omega t) \Theta(t) T^x_\alpha.
\label{eq:rtls}
\eeq
Here, we have used the result (from Sec.~\ref{restrans}) that a RTLS with splitting $\varepsilon$ is typically one of size $x(\varepsilon) \sim \zeta \log(W/\varepsilon)$, and that the corresponding dipole matrix element of the electric field is $A x \sim A \zeta \log(W/\varepsilon)$. The density of these RTLS's is also given by similar reasoning. The number of available states at scale $x$ goes as $\exp(sx)$, and the corresponding many-body level spacing is $W x \exp(-s x)$ [since $W x$ is the energy bandwidth of a region of size $x$]. Substituting $x(\varepsilon)$ into this expression, we immediately arrive at the result

\beq
\rho(\varepsilon) \sim \varepsilon^{-s\zeta}
\eeq
Note that this is the density of states of RTLS's, \emph{not} necessarily that of effective $\tau$ spins. 

With these assumptions, we can apply the rotating wave approximation to \eq{eq:rtls} and the dynamics of RTLS's becomes exactly solvable. In what follows we shall further simplify by neglecting the logarithmic correction due to the dipole moment. Now one can use the Rabi formula to find that at time $t$, the energy variance of a single RTLS is given by

\beq
(\Delta E_\alpha)^2 \simeq \frac{4A^2 \zeta^2 \varepsilon_\alpha^2}{\Omega_i^2}  \sin^2 (\Omega_\alpha t)
\eeq
where $\Omega_\alpha \equiv \sqrt{(A\zeta)^2 + (|\varepsilon_\alpha| - |\omega|)^2}$ is the Rabi frequency of RTLS $\alpha$.

\begin{figure}
\begin{center}
  \includegraphics[width=0.48\textwidth]{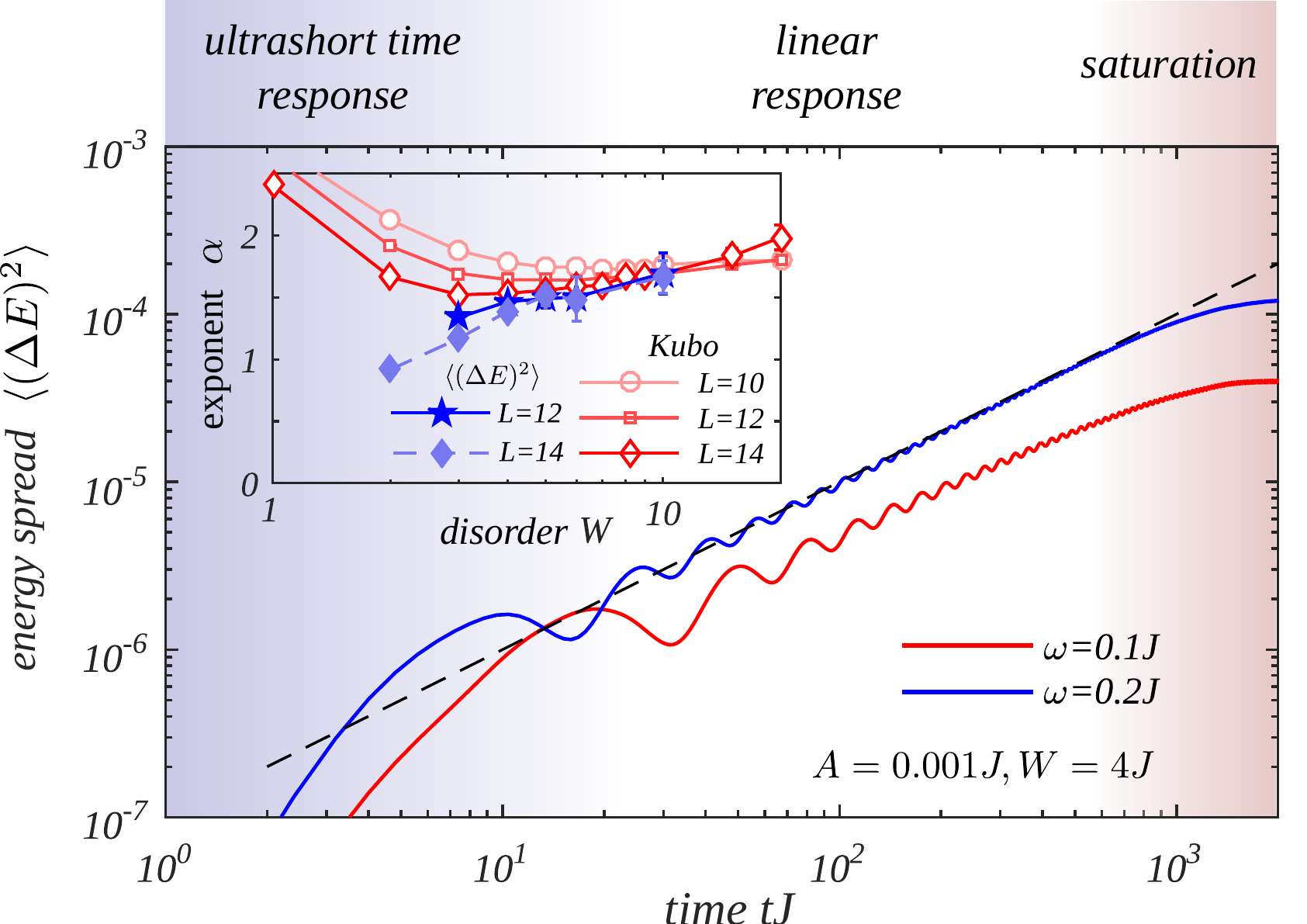}
  \caption{\textbf{Energy spread in the linear response regime.} Time evolution of the energy spread calculated numerically for a driven and disordered Heisenberg model, \eq{eq:h}, of size $L=12$ with open boundary conditions, driving amplitude $A=0.001J$, disorder $W=4J$, and two different driving frequencies $\omega=0.1J$ and $\omega=0.2J$. In a considerably large time regime, the energy spread is linear, as indicated by the reference curve with linear slope, dashed line. The linear response regime separates the ultrashort time regime, $ t\lesssim 1/\omega$, in which the driving frequency cannot be resolved from the regime in which the TLS are saturated, $t \gtrsim 1/A$. Inset: Exponent $\alpha$ obtained from the scaling energy spread  with the drive frequency $\langle (\Delta E)^2 \rangle \sim A^2 \omega^{\alpha}t$, blue stars and diamonds, compared to the exponent of the optical conductivity obtained directly from the Kubo formula~\cite{mbmott}.  }
\label{fig:linResponse}
\end{center}
\end{figure}

To get the response of the full system, one ensemble-averages the response of the RTLS's. This gives the expression

\beq\label{into}
(\Delta E)^2(t) = W^{s \zeta - 1} \int d\varepsilon \left[ \frac{4 A^2 \varepsilon^{2-s \zeta}}{\Omega^2} \sin^2 (\Omega t)\right].
\eeq
This integral has four regimes. At short times compared with $1/W$ 
it goes as $A^2 t^2$. At long times compared with $t \agt 1/A \agt 1/\Omega$, it saturates. There are two intermediate regimes: $1/W \ll t \ll 1/\omega$ and $1/\omega \ll t \ll 1/A$. The former regime is not of interest to us: at these timescales, the frequency $\omega$ cannot be resolved. Thus, we can specialize to $1/\omega \ll t \ll 1/A$. Here, the integral \eqref{into} splits into three parts: from 0 to $\omega - 1/t$, from $\omega - 1/t$ to $\omega + 1/t$, and from $\omega + 1/t$ to $W$. In the ``outer'' regimes, we can approximate $\sin^2 x \simeq 1/2$, and in the ``inner'' regime, we can expand it as $\sin^2 x \simeq x^2$. Using these results, we find that the leading $t$-dependence in this regime is given by 

\beq\label{lrfinal}
 (\Delta E)^2_\text{LR}(t) \sim W^{s \zeta-1} A^2 \omega^{2 - s \zeta} t
\eeq
This is, as expected, proportional to the linear response conductivity $\sigma(\omega) \sim \omega^{2-s\zeta} \equiv \omega^\alpha$~\cite{mbmott}, \textit{cf.} \fig{fig:linResponse}.

\subsection{Linear response in the thermal phase}\label{lrth}

We now turn to the thermal phase, and briefly consider how linear response emerges there. As we shall eventually be interested in finite-size thermal blocks in the insulating phase, we focus on a finite thermal system of size $L$, with an intrinsic thermalization time $\sim 1/W \ll 1/\omega$. Using the ``off-diagonal'' part of the eigenstate thermalization hypothesis~\cite{srednicki1994}, one can estimate the matrix elements of the electric field between many-body eigenstates of the thermal system as $M(L) \sim A \exp(-sL/2)$. (Here, we are ignoring subleading factors of $L$ that are not in the exponent.) When $L$ is large enough, such matrix elements are always much smaller than $\omega$; hence resonant transitions always dominate Landau-Zener transitions, and the Golden Rule is appropriate. Furthermore, the timescale at which linear response breaks down due to saturation in a large, deeply thermal inclusion is \emph{not} determined by $t \simeq 1/A$, as in the previous section. Instead, it is set by the shorter of the following two timescales. (1)~The timescale on which the occupation of the initial eigenstate is appreciably depleted. This timescale is set by the Golden-Rule rate $\sim A^2/W$, which is independent of $L$ but is parametrically longer than in the localized phase, since $A/W \ll 1$ for our purposes. (2)~The timescale on which a particular final state has an appreciable chance of being populated. This is set by the matrix element $M(L) \sim A \exp(-sL/2)$. For very large thermal systems, process (1) governs saturation, whereas for small thermal systems (such as some of the Griffiths regions we will consider below), process (2) governs saturation. A crossover between these processes takes place when $L \sim (2/s) \log(W/A)$. 

\section{Regimes of nonlinear response\label{se:intdrive}}

The perturbative resonances discussed in the previous section saturate on a timescale $\sim 1/A$. When $A \agt \omega$, these resonances are essentially saturated within the first drive cycle. Thus, any heating that occurs after the first drive cycle is due to slower, more collective processes. We now consider various types of such processes: (A) perturbative resonances that are slow compared with the main Mott transitions, and thus give rise to slower heating; (B) perturbative resonances that are higher-order in the drive amplitude; (C) thermal Griffiths inclusions; and (D) Landau-Zener transitions. 

\subsection{Anomalously distant resonant pairs\label{se:subMott}}

First, we extend the analysis of Sec.~\ref{lr} to times that are long compared with $1/A$; at these times, the dominant Mott resonances have saturated. However, rare Mott pairs with anomalously small Rabi frequency still exist, as do pairs of states with splitting $\omega$ at larger scales than $x_\text{Mott}$. We expect the latter to dominate, as they are more abundant, so we shall focus on them. Unlike the Mott pairs, these subleading resonances are \emph{induced} by the drive: i.e., although the pairs are split by $\omega$, this splitting is due to detuning rather than hybridization. Thus they are hybridized by the \emph{off-diagonal} matrix elements of the drive. The hybridization is given (at distances $x \gg x_{\text{Mott}}$) by $\tilde{A}(x) \sim A x \exp(-x/\zeta)$.  Moreover the number of these resonances increases with distance as $\exp(s x)$. Now let us consider the dynamics on a timescale $t$. On this timescale, resonances with $\tilde{A}(x) \agt 1/t$ will have saturated and do not contribute any further to heating. However, further-out drive-induced resonant pairs will still be absorbing linearly. The absorption at time $t$ is thus dominated by resonances with $\tilde{A}(x) \simeq 1/t$. Plugging this into Eq.~\eqref{lrfinal}, the contribution from these resonances to heating goes as 

\begin{equation}
   (\Delta E)_\text{Anom.-Mott}^2(t) \sim \tilde{A}(x)^2 e^{sx} \omega^2 t/W \sim A^{s\zeta} \omega^2 t^{s\zeta - 1} / W.
\end{equation}
Stability of the MBL phase entails $s\zeta < 1$, so that these processes give rise to a slow, power-law approach to saturation on timescales $t \agt 1/A$ even when the drive is weak.

\subsection{Resonances from higher-order processes in the drive\label{se:hop}}

In the previous sections we considered one way in which the drive can induce $n$-particle rearrangements: namely, that the expansion of the electric field in terms of effective spins has matrix elements for rearranging $n$ spins. For large $n$ such a process is suppressed because it falls off as $\exp(-n/\zeta)$ [see Sec.~\ref{model}]. Nevertheless, it is still \emph{leading-order} in the drive amplitude $A$. When the drive amplitude is large, one must also consider resonant $n$-particle rearrangements that are higher-order in the drive amplitude. For instance, one can rearrange $n$ effective spins by going to $n$th order in the drive. The amplitude for such a process can be estimated in perturbation theory as $\tilde A_n \sim A^n / W^{n - 1}$ (up to a combinatorial factor) because the typical energy change upon flipping an effective spin is $W$. To see which type of $n$ spin rearrangement is more important, one must compare $\zeta$ with $1/\log(W/A)$; the bigger of these will dominate. We have considered the former type (first-order in $A$) above; now we consider the latter (high-order in $A$).

The resonances that go as $A^n$ saturate only on timescales $t \sim 1/\tilde A_n$; thus $n$th order processes can dominate response once all lower-order processes have saturated. At a time $t$, the dominant unsaturated resonances are of order $n$ such that $\tilde A_n \sim 1/t$, and thus $n(t) \sim \log t / \log(W/A)$. These $n$th order processes can be analyzed in terms of the Rabi formula, precisely as in Sec.~\ref{lr} but replacing $A$ with the renormalized Rabi frequency $\tilde A_n \sim 1/t$. Thus, $(\Delta E)_n^2 \sim \tilde A_n^2 e^{sn} t $. Substituting for $\tilde A_n$ and $n(t)$, we arrive at the result

\beq
(\Delta E)_\text{Hi.-Res.}^2 \sim t^{-1 + \mathrm{const.} s /[ \log(W/A) ]}
\eeq
up to an overall constant due to the combinatorics of $n$th order processes. Thus, higher-order processes give rise to a power law that is (a)~sensitive to the drive amplitude $A$, and (b)~can be either positive or negative. In the limit $A, \omega \rightarrow 0$, we expect these processes to be subleading (since $A/W \rightarrow 0$) but for numerically accessible frequencies, it is plausible that these processes will be relevant for the late-time dynamics.

\subsection{Thermal Griffiths inclusions} \label{thgriff}

So far, we have focused on heating processes involving isolated two-level systems inside the MBL phase. A separate channel for response and heating comes from thermal ``inclusions,'' or thermalizing islands embedded in a localized bulk. We expect this channel to be particularly important near the delocalization transition. 
To explore it, we first discuss the response due to a single deeply thermal segment of length $L$, with linear-response conductivity $\sigma_{th}(\omega)$. As discussed in Sec.~\ref{lrth}, the heating rate of this inclusion is given by the linear-response result $\sim A^2/W \sigma_{th}(\omega)$, and saturates on a timescale $t_s \simeq 1/A \min(e^{sL/2}, W/A)$. We are interested in relatively small islands, and in the $A/W \rightarrow 0$ limit, so we shall consider the first case, $t_s \sim (A e^{-sL/2})^{-1}$. Moreover, the probability of having a thermal inclusion of length $L$ goes as $p^{L}$, where $p$ is some probability per unit length that vanishes deep in the localized phase, and presumably approaches unity at the delocalization transition~\cite{Agarwal, mbmott}.

Let us now consider the response at time $t$, such that $1/A \alt t \alt W/A^2$. At this time, the smallest Griffiths regions that have not saturated have size $L \simeq (2/s) \log(A t)$; the density of such rare regions decreases as $t^{-2\log(1/p)/s}$, and each region contributes $(A^2/W) \sigma(\omega) t$ to the energy spread.
%
Combining these results, we find that the Griffiths contribution (from strongly thermal inclusions) to the heating rate is given by 

\beq\label{thinc}
(\Delta E)^2_{\text{Griff}}  \sim A^{2 - g} t^{1 - g}.
\eeq
where $g \equiv 2\log(1/p)/s$ is expected to be generically small, and thus the overall exponent is expected to be generically positive, near the MBL transition. 

The above estimate is for the contribution from \emph{thermal} inclusions. However, it is possible that fractal \emph{critical} inclusions could give an even faster heating rate: in particular, it seems that the probability of critical inclusions might vanish as $\exp(-g' L^{d_f})$, where $d_f < 1$~\cite{trithep}. This might lead to a parametrically faster energy spread than the simple thermal inclusions we are considering: however, at present the heating behavior of such critical inclusions is unclear. 


\subsection{Landau-Zener transitions\label{se:discLZ}}

In addition to the perturbative resonances discussed above, one expects that absorption due to Landau-Zener processes should also be important in the low-frequency limit. The Landau-Zener contribution has two regimes, depending on the scale $x_\text{ad}$, which separates mostly adiabatic resonances from mostly diabatic ones: when $x_\text{ad} \ll x_\text{LZ}$ [i.e., $A^{1-s \zeta/2} \alt \omega^{s \zeta/2} W^{1 - s \zeta}$] the ``active'' Landau-Zener transitions---i.e., those that have an appreciable probability of being both diabatic and adiabatic---are rare and isolated, and can thus be treated individually. In the opposite limit $x_\text{ad} \gg x_\text{LZ}$, the Landau-Zener transitions form a percolating network, and the system delocalizes. 

\subsubsection{Isolated Landau-Zener transitions}

We now estimate the heating rate due to isolated Landau-Zener transitions. In general, a transition that is always adiabatic or always diabatic does not contribute to energy spread (see Sec.~\ref{introlz}); rather, the timescale on which a given Landau-Zener transition acts dissipatively (or, equivalently, loses memory of its initial state) is given by 

\beq\label{TX}
T(x) \simeq 1/[\omega P_\text{ad}(x) (1-P_\text{ad}(x))].
\eeq
In the regime we are considering, Landau-Zener transitions are isolated. Thus on timescales long compared with $T(x)$, all Landau-Zener transitions at a length-scale $x$ are saturated and do not contribute to heating. Let us consider the response at time $t$. Then, the leading contribution to heating will be from transitions with $T(x) \simeq t$. At long times, this means the transitions that have not yet saturated are mostly adiabatic or mostly diabatic. The phase space for mostly diabatic transitions, $x\gtrsim x_\text{ad.}$, is larger (because these correspond to larger-scale rearrangements, of which there are more) so we focus on those. For such transitions, $P_{\text{ad}} \ll 1$, so we can simplify~\eq{TX} by approximating $P_{\text{ad}} \sim W^2 \exp(-2x/\zeta)/(A x \omega)$ to write
\beq
x(t) \simeq (\zeta/2) \log[W^2 t/(A x)].
\eeq
At a length-scale $x(t)$, there are $(A/W) \exp[sx(t)] \sim A^{1 - s\zeta/2} t^{s \zeta / 2}$ Landau-Zener crossings. Each of these contributes $\sim A x(t)$ of energy. Thus, up to logarithmic factors, the total Landau-Zener contribution to energy spread is

\begin{figure*}
\begin{center}
  \includegraphics[width=0.98\textwidth]{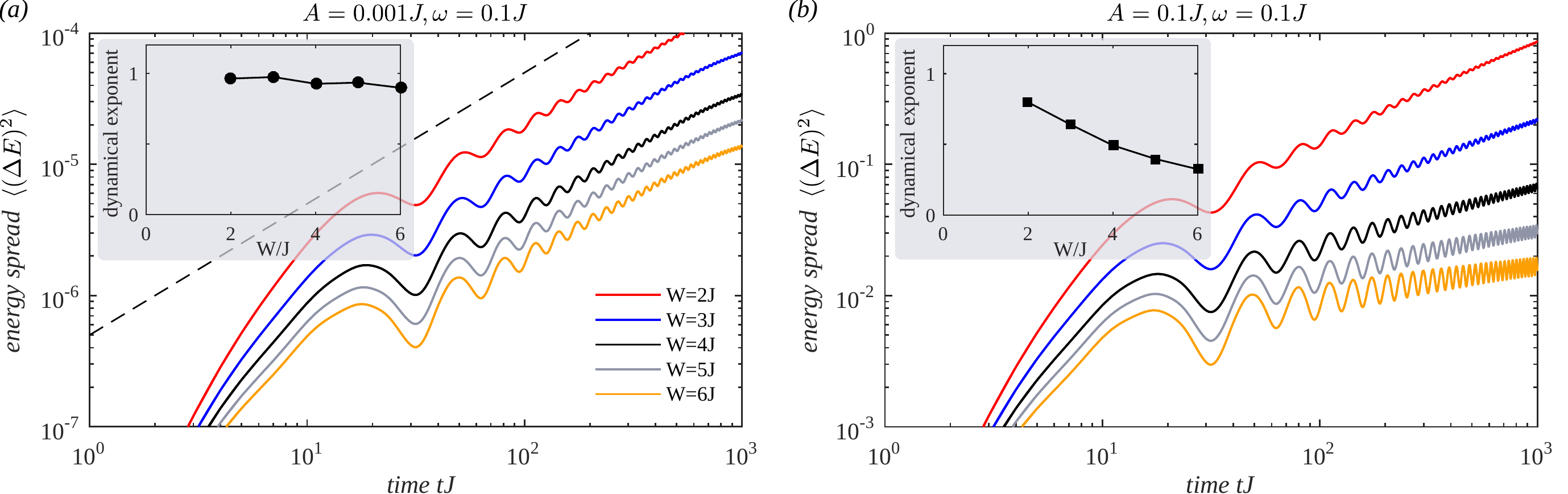}
  \caption{\textbf{Entering the non-linear regime}. \fc{a} In the  regime of weak drive $A=0.001J$ the energy spread $\langle (\Delta E)^2 \rangle(t)$ is linear in a large time window irrespective of the disorder strength, while in the strong drive regime, \fc{b}, the response is sub-linear with an exponent that strongly decreases with increasing disorder strength, see insets for the respective dynamical exponents extracted from a powerlaw fit to the data. In all cases the driving frequency $\omega=0.1J$ and systems of size $L=12$ with open boundary conditions have been used. The dashed line in \fc{a} indicates a growth that is linear in time. }
\label{fig:en_W}
\end{center}
\end{figure*}
\beq
(\Delta E)^2_\text{LZ} \sim A^{2 - s \zeta/2} t^{s \zeta/2}
\eeq
This analysis is incomplete because it ignores interference between subsequent Landau-Zener transitions. Thus, it would naively suggest that any degree of freedom always delocalizes at sufficiently long times, because a Landau-Zener crossing inevitably takes place. Interference effects qualitatively modify this picture at long distances, as discussed in App.~\ref{app:LZ}, ensuring the stability of the MBL phase. 
%
In a simple model where the Landau-Zener transitions can be treated as entirely isolated, this approach gives us the late-time asymptotic result is $(\Delta E)^2_{\mathrm{LZ, int}} \sim A^{2 - (s \zeta/2)} \omega^{s \zeta/2} t^{s\zeta - 1}$.
It is not clear, however, that this result is correct for the setup we have in mind, in which the drive is suddenly turned on at time $t = 0$. In this setup, even the \emph{typical} effective spins (which are not involved in resonant or Landau-Zener transitions) nevertheless exhibit weak precessional dynamics and only undergo quantum revivals at very long times (as discussed, e.g., in Ref.~\cite{vpm}). Thus, the environment of a given Landau-Zener transition is never exactly $\omega$-periodic, which complicates a full analysis of interference between Landau-Zener transitions. 

\subsubsection{Percolating network of Landau-Zener transitions}

When the drive amplitude is large enough that $x_\text{ad} \approx x_\text{LZ}$, then a chain of Landau-Zener transitions percolates through the system. This leads to a delocalized steady state in which the system heats up to infinite temperature. It seems plausible (as discussed below) that the delocalized state near the percolation transition exhibits anomalous transport~\footnote{Note that although energy is not conserved in the driven system, there might be other conserved quantities (such as $\sigma^z$ in the XXZ model we studied numerically), so that it is meaningful to discuss transport.}. Even in such an anomalous-transport phase, however, the long-time heating behavior, for finite-frequency driving, is expected to be \emph{linear} in time: i.e., the finite-frequency linear response coefficients are well-defined in this phase in the high-temperature limit [see Ref.~\cite{bouchaud_rmp}, Sec. 5.4]. Nevertheless, close to the drive-induced delocalization transition, the typical relaxation timescales are very long; absorption on much shorter timescales is dominated by single Landau-Zener transitions, as discussed in the previous section.
We emphasize that this ``physical'' charge diffusion is not to be confused with the ``fictitious'' diffusion process discussed in Sec.~\ref{se:dynresp}. 

\subsection{Summary and Floquet perspective}

In this section, we have discussed various mechanisms that cause heating on timescales $t\agt 1/A$---anomalously large-scale (and therefore slow) Mott resonances, higher-order processes in the drive amplitude, thermal Griffiths inclusions, and Landau-Zener transitions. We have argued that all these effects give rise to nonlinear heating characterized by continuously varying power-laws in time (owing to a wide distribution of saturation timescales), but the exponent can be negative, e.g., with anomalously large Mott resonances, or positive, as with Griffiths inclusions and Landau-Zener crossings (in the intermediate-time window where interference effects are not important). 

These results are relevant for intermediate times. However, at asymptotically late times, these behaviors all reduce to two types: power-law approach to a saturated value as $(\Delta E)^2_\infty - (\Delta E)^2 (t) \simeq t^{-\phi}$, or linear growth in case the Landau-Zener transitions percolate. These can be understood from the following complementary perspective. One can regard the protocol we have discussed as being a quantum quench into an effective Floquet Hamiltonian $\hat H_F$, defined via $\exp[-i2 \pi\hat H_F/\omega] \equiv \hat U(2 \pi/\omega)$, which is itself either MBL or delocalized. The late-time behavior after such quenches is well understood in both the MBL and thermal phases. When the Floquet Hamiltonian is itself localized, local operators approach their eventual expectation values with a slow power law~\cite{mbmott, serbyn_quench}. On the other hand, when the Floquet Hamiltonian is deep in its thermal phase, one \mkch{naively} expects essentially linear heating at times $\gg 1/A$.

This Floquet perspective also suggests that near the drive-induced delocalization transition, the system should be in a Griffiths phase with anomalous charge diffusion and associated slow dynamics. The delocalization transition point depends on $\zeta$, which is spatially fluctuating. Thus, in the delocalized phase near the transition, there will be regions of the system, e.g., with anomalously small $\zeta$, that are locally still in the Floquet-MBL phase, and these will presumably act as transport bottlenecks. Late-time dynamics after a quench into such a Floquet Hamiltonian with \emph{anomalous charge diffusion} (note that charge, unlike energy, is conserved by the drive) has not been explored in detail. A simple estimate is that the heating at late times $t$ is governed by the density of locally insulating regions at time $t$ (as these take a long time to heat up). This would then suggest~\cite{gopalakrishnan_griffiths_2015} that the late-time approach to saturation should go as $t^{-1/z}$, and thus should go logarithmically at the critical point. This is consistent with what is seen numerically (see below, and Ref.~\cite{Pollmann}). 


\section{Numerical results}\label{se:num}
To support our analytic estimates, we perform numerical simulations on the random-field XXZ chain, described by the Hamiltonian
\begin{equation}
 \hat H = \frac{J}{2} \sum_{\langle ij \rangle} (\hat S_i^+ \hat S_{j}^- + \text{h.c.}) + J_z \sum_{i} \hat S_i^z \hat S_{i+1}^z  +  \sum_i h_i  \hat S_i^z\;,
 \label{eq:h}
\end{equation}
where $h_i$ is a local random field drawn from a uniform distribution of range $[-W,\,W]$, $J$ is the spin exchange scale, and $J_z$ the spin-spin coupling strength, which we set equal and use as energy unit throughout this work. The monochromatic drive
\begin{equation}
 \hat H_\text{drv.}(t) = A \sin \omega t \, \sum_i x_i \hat S_i^z
\end{equation}
is switched on for $t \geq 0$. For our purposes it is necessary to use a \emph{monochromatic} drive, instead of the square-wave drives in Refs. \cite{ldm, ponte_periodically_2015}. While implementing a square-wave drive is numerically simpler, it complicates the extraction of frequency-dependent response, because the higher harmonics of the drive (corresponding to larger $\omega$) have higher conductivity and thus dominate the heating at short to intermediate times.  We initialize the dynamics by an eigenstate of $\hat H$ and propagate it in time by discretizing the time evolution operator $\hat U(t)= \mathcal{T}_t \exp [-i \int_0^t dt' (\hat H + \hat H_\text{drv.}(t'))] \approx \prod_{n=1}^N \exp[-i \Delta t (\hat H + \hat H_\text{drv.}(n \Delta t))]$, where $\Delta t=t/N$. The stepwise propagation is performed by Lanczos time evolution which allows us to efficiently update the instantaneous Hamiltonian $\hat H + \hat H_\text{drv.}(n \Delta t)$. All data is taken for systems with open boundary conditions in order to avoid the jump of the electric field in space. 

\begin{figure}
\begin{center}
  \includegraphics[width=0.48\textwidth]{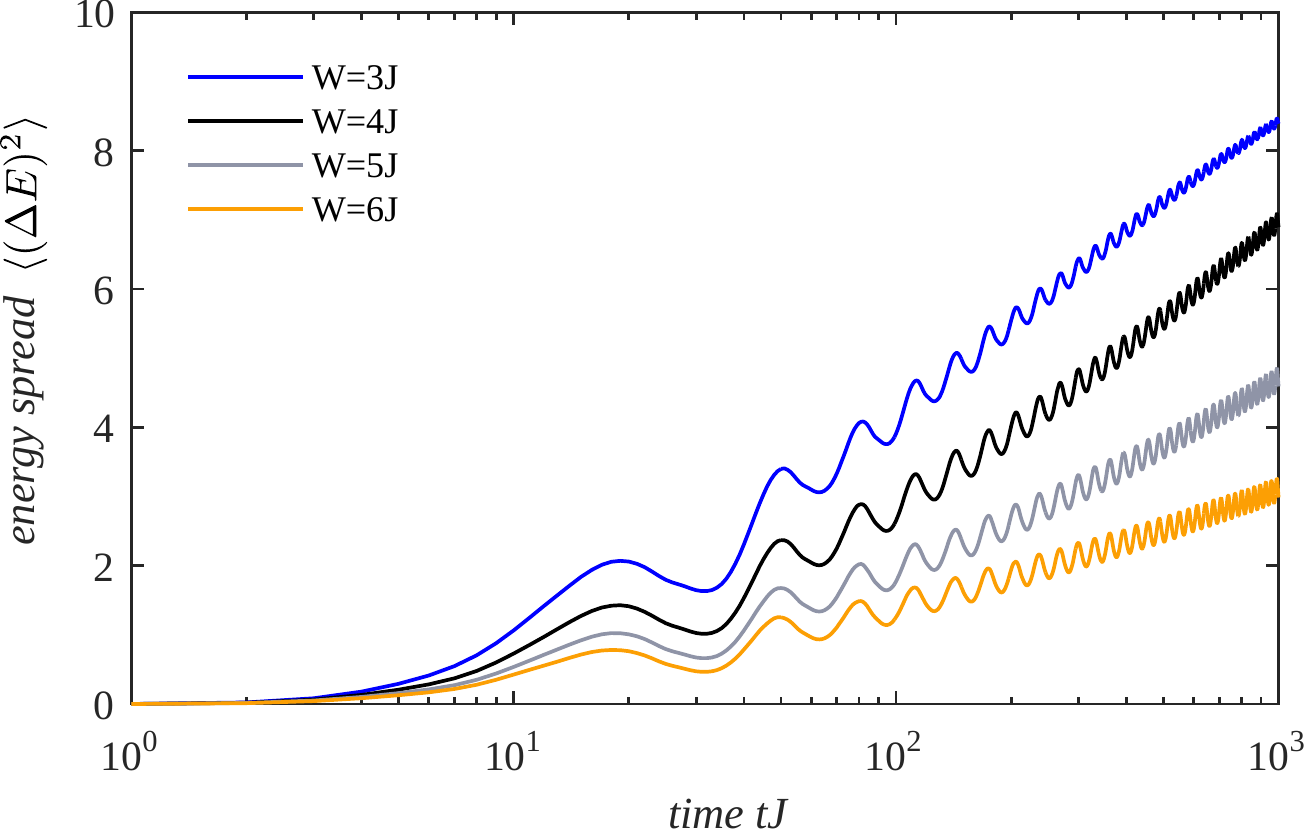}
  \caption{\textbf{Response of percolating Landau-Zener transitions}. In the strong drive limit, Landau-Zener transitions form a percolating network and the effective Floquet Hamiltonian is delocalized. In the \mkch{crossover to that regime, we find that the energy spread grows slower than naively expected as a logarithm} in time (in contrast to the powerlaw growth which we predict in the linear-response and the intermediate non-linear regime), consistent with the findings of Ref.~\cite{Kozarzewski,Pollmann}. The data is shown for driving frequency $\omega=0.1J$, drive amplitude $A=J$, and system size $L=12$, for different values of the disorder strength $W$ as stated in the legend. }
\label{fig:en_deloc}
\end{center}
\end{figure}
\begin{figure*}
\begin{center}
  \includegraphics[width=0.98\textwidth]{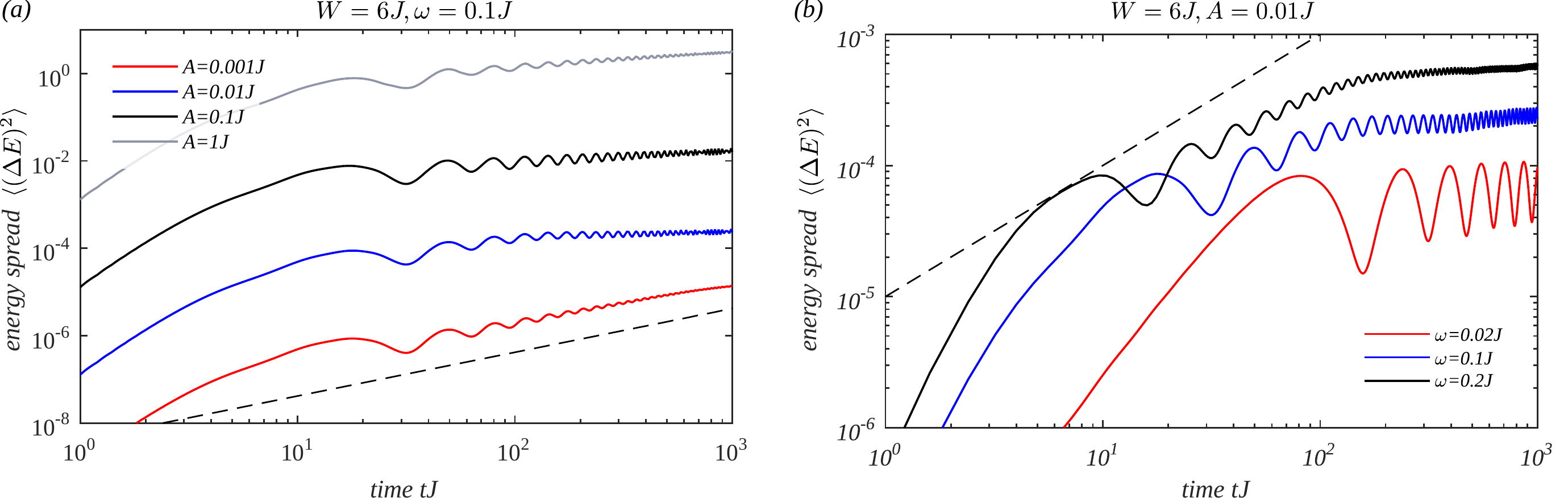}
  \caption{\textbf{Amplitude and frequency dependence of the energy spread}. Energy spread in time $\langle (\Delta E)^2 \rangle(t)$ for \fc{a} different values of the driving amplitude $A$ and fixed frequency $\omega=0.1J$ and \fc{b} driving amplitude $A=0.01J$ and different values of the frequency $\omega$. In both cases the disorder is $W=6J$ and the system size $L=12$. The dashed line indicates a linear slope, which corresponds to linear response. Linear response occurs at relatively low amplitudes and large frequencies: thus, in panel (a), a linear slope is evident only for small drive amplitude; in panel (b), for fixed amplitude, a linear-response regime \emph{emerges} as the drive frequency is increased (black curve). The latter behavior is a distinctive feature of the MBL phase.}
\label{fig:en_A}
\end{center}
\end{figure*}

\subsection{Linear response regime}

First, we check for the validity of linear response theory, which should apply for any fixed frequency when the amplitude goes to zero. We find, indeed, that for small-amplitude driving there is a considerable regime where the energy spread is linear (as linear-response theory would predict), see \fig{fig:linResponse}. In this regime, increasing the drive strength does not change the \emph{exponent}, but causes saturation to set in sooner. 

In order to further benchmark this dynamical regime against linear response theory, we extract the rate of energy spread (i.e., the prefactor of the linear regime) as a function of frequency, and compare it with the linear response exponents obtained in Ref. \cite{mbmott} for systems with open boundary conditions. We find that the two sets of exponents are largely consistent, see inset of \fig{fig:linResponse}. Near the MBL transition and on the ergodic side a direct comparison gets complicated by finite size effects which are different for the Kubo conductivity and the energy spread. However, deep in the localized phase, the exponents agree reasonably.

\subsection{Nonlinear response and amplitude dependence}

We now turn to the nonlinear response of larger drive amplitudes. In \fig{fig:en_W} we show the energy spread for different values of disorder strength $W$, ranging from the ergodic to the localized phase for fixed driving frequency $\omega=0.1J$ and drive strength $A$ that is weak in \fc{a} $A=0.001J$ and strong in \fc{b} $A=0.1J$. In the weak drive limit the response is linear in a wide time window irrespective of the disorder strength $W$, \textit{cf.} inset which shows the exponent as a function of disorder strength. By contrast, for strong drive, the powerlaw exponent of the energy spread decreases significantly with disorder strength (inset), indicating that the sublinear regime has been entered. This behavior has been predicted by all the mechanisms discussed in Sec. \ref{se:intdrive}. For even stronger drive, a percolating network of Landau-Zener transition forms, and the energy spread changes from power-law slow heating to logarithmically slow heating, \fig{fig:en_deloc}, consistent with the findings of Ref.~\cite{Kozarzewski,Pollmann}. \mkch{This logarithmic growth of the energy spread in time is characteristic for the crossover regime to the thermal phase~\cite{Pollmann} and is slower than the naively expected linear growth for a Floquet Hamiltonian being deep in the thermal phase.}

We now study the drive amplitude dependence at strong disorder $W=6J$, \figc{fig:en_A}{a}. For intermediate driving amplitudes $A\gtrsim 0.01J$, the energy spread grows sublinearly in time, with a power law that increases weakly with the amplitude. We conjecture that this dependence on the drive amplitude arises from higher oder resonances as discussed in Sec.~\ref{se:hop}.
The frequency dependence of the energy spread $\langle (\Delta E)^2\rangle$ for intermediate driving amplitude $A=0.01J$ transitions from sublinear growth at low driving frequency to an intermediate linear growth at higher frequency, see \figc{fig:en_A}{b}, in agreement with the picture of saturating two-level systems.

\begin{figure*}
\begin{center}
  \includegraphics[width=0.98\textwidth]{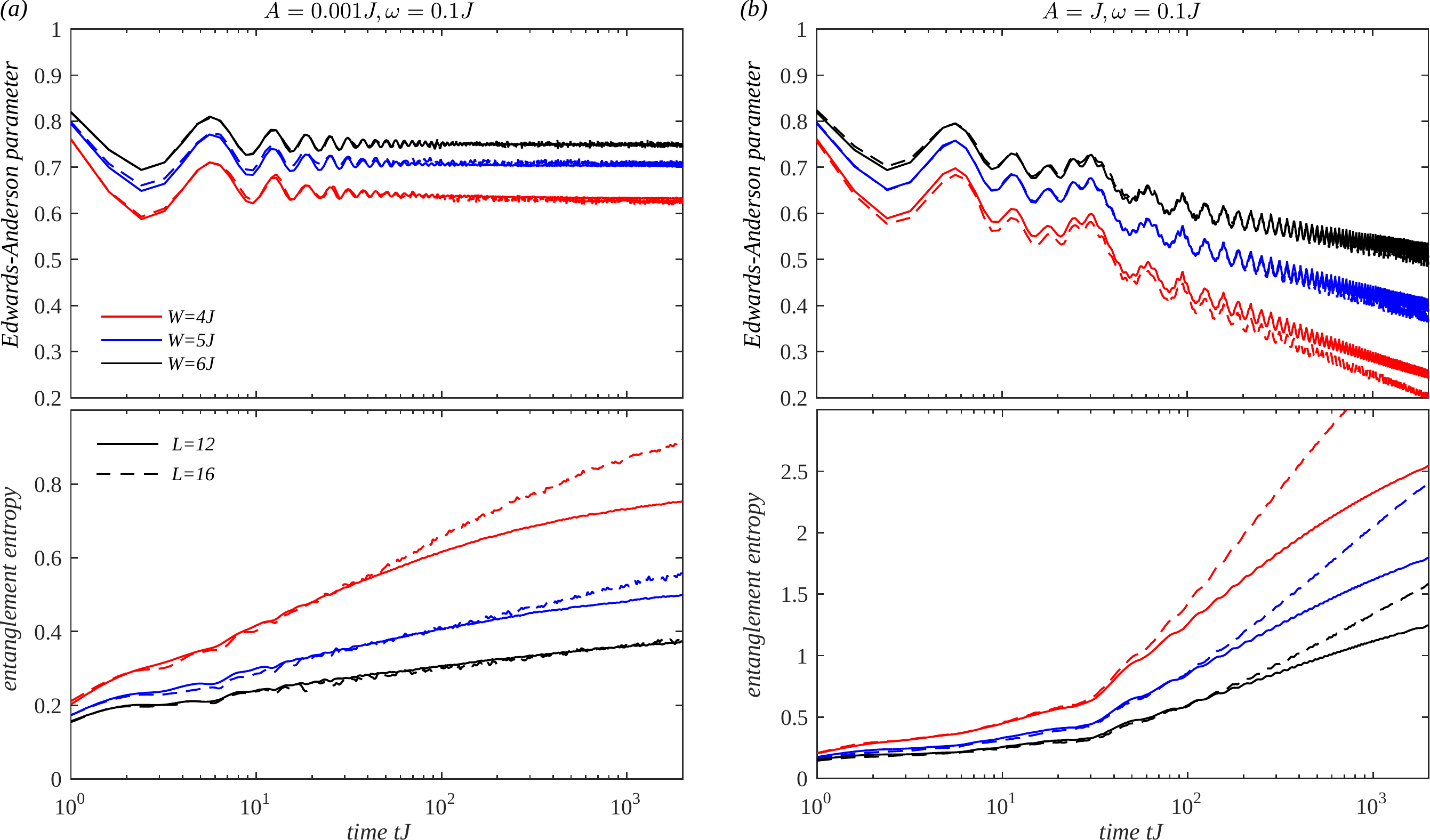}
  \caption{\textbf{Edwards-Anderson order parameter and von Neumann entanglement entropy}. The Edwards-Anderson order parameter (or Hamming distance), top row, and entropy, bottom row, for \fc{a} weak drive $A=0.001J$ and \fc{b} strong drive $A=J$, driving frequency $\omega = 0.1 J$, and three values of disorder strength $W=\{4,5,6\}J$. The system size is $L=12$, solid lines, and $L=16$, dashed lines. For weak drive, the system and hence the effective Floquet Hamiltonian remains localized, while for strong drive, it delocalizes manifesting in a decay in the Edwards-Anderson parameter and a strong increase of the entanglement entropy. }
\label{fig:pol}
\end{center}
\end{figure*}

\subsection{Additional probe: Edwards-Anderson parameter and von Neumann entanglement entropy}

A complementary perspective to switching on the periodic modulation is to regard it as a \emph{quantum quench} from the original Hamiltonian to the Floquet Hamiltonian. From this perspective, a key question is whether the corresponding Floquet Hamiltonian is localized or delocalized. We have explored this issue by looking at the evolution of the Edwards-Anderson parameter (or Hamming distance~\cite{hauke_many-body_2015})
\begin{equation}
 \chi(t) = \frac{4}{L} \sum_i \bra{p} \hat U^\dag(t) S_i^z \hat U(t) S_i^z \ket{p} ,
\end{equation}
where $\ket{p}$ is an arbitrary product state which we take as a random initial state. A special case of the Edwards-Anderson order parameter is the decay of contrast of an initial staggered magnetization, which has been used as an order parameter in recent experiments~\cite{schreiber2015observation, smith2015, bordia2015}. In the MBL phase and for a drive in linear-response regime $A\ll\omega$, the Edwards-Anderson order parameter saturates in the infinite time limit to a finite value, since at weak drive the effective Floquet Hamiltonian remains to be localized, \figc{fig:pol}{a}, top. By contrast, in the strong drive limit $A\gg\omega$, it decays to zero, since the effective Floquet Hamiltonian is thermal, \figc{fig:pol}{b}, top, which confirms that for the strong drive considered in \fig{fig:en_deloc} a percolating network of Landau-Zener transitions has been formed. 

In addition, we have computed the von Neumann entanglement entropy growth due to the drive, \fig{fig:pol} bottom row. Well in the localized regime, $W=6J$ and for weak driving amplitude  $A=0.001J$, the entanglement entropy does not exhibit any finite size effects, as the effective localization length of the Floquet Hamiltonian $\hat H_F$ is expected to be much smaller than the system size, while closer to the transition $W=\{4J, 5J\}$ the simulated system sizes are too small for the entanglement entropy to refrain from finite size effects. By contrast, for strong drive $A=J$, a substantial system size dependence is observed for all values of the disorder strength, which also confirms the delocalized nature of the effective Floquet Hamiltonian. In the strong drive limit, the entanglement entropy starts to grow abruptly after approximately half of a driving cycle $t \sim \pi/\omega$, corresponding to the Landau-Zener crossing time scale.

\section{Discussion}\label{se:dis}

\begin{figure}[b]
\begin{center}
\includegraphics[width = 3.4in]{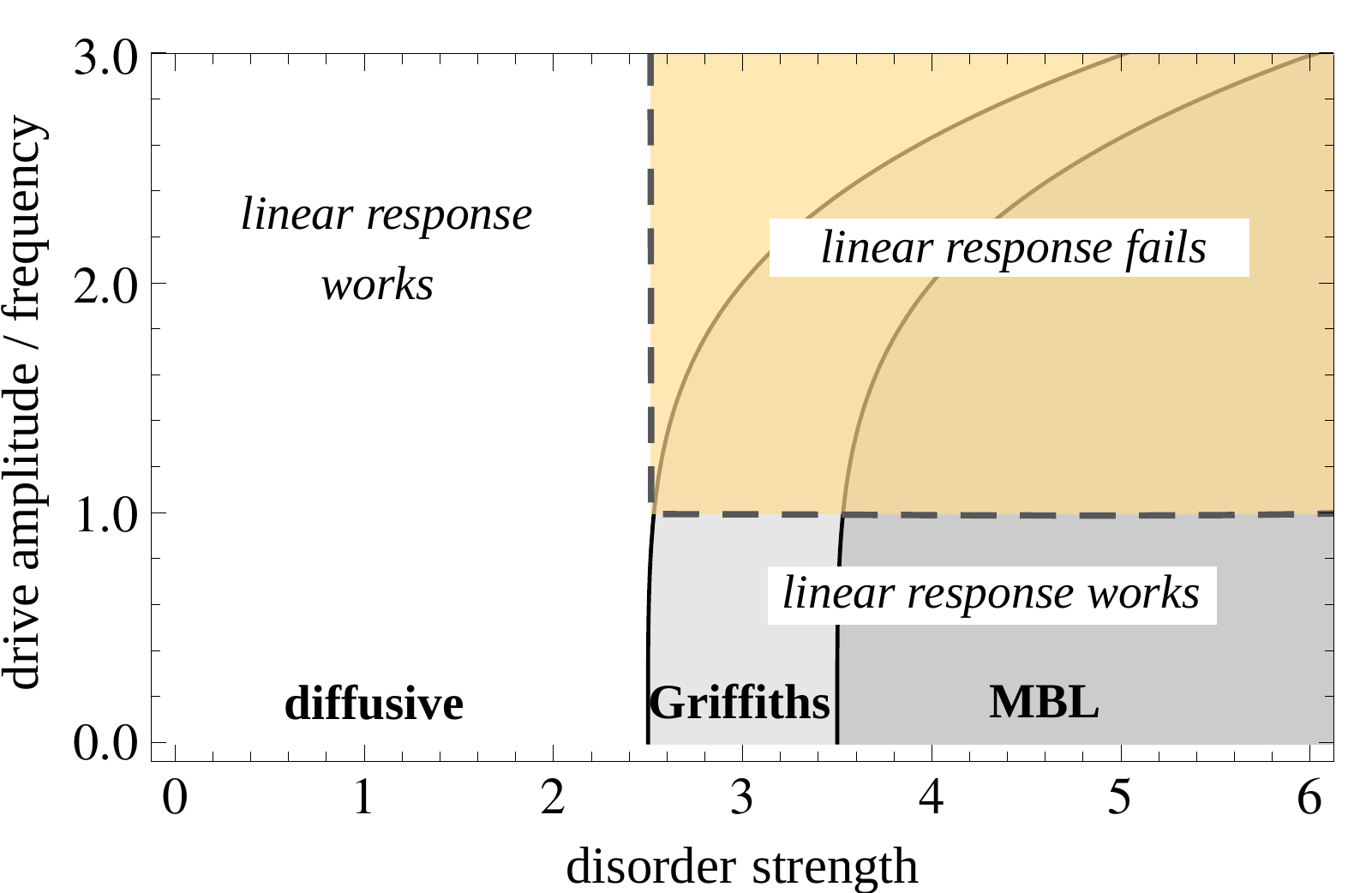}
\caption{Schematic phase diagram showing regimes of dynamical response in a driven disordered system, at a fixed drive frequency, as a function of drive amplitude and disorder strength. Both frequency and amplitude are taken to be small relative to the characteristic energy scales (e.g., bandwidths) of the system. Solid lines indicate steady-state transitions between a diffusive thermal phase, a subdiffusive Griffiths phase and an MBL phase. The MBL phase is destabilized as the drive amplitude increases~\cite{ldm, abanin2014theory}. Dashed curve shows the crossover between linear and nonlinear response \emph{in the transient dynamics} of a driven system. This crossover is in general distinct from the steady-state phase transitions.}
\label{pd}
\end{center}
\end{figure}

Our objective in this work was to identify regimes for which linear response theory correctly predicts the dynamics of a driven MBL system, and those for which the response is essentially nonlinear. Our key results are that heating in the finite-frequency, weak-drive regime is essentially conventional (corresponding to linear response theory with the appropriate conductivity), whereas the behavior at larger drive amplitudes (or lower frequencies) is not. It seems that in this regime neither the amplitude-dependence nor the time-dependence of the heating correspond to linear response predictions. Rather, as we discussed, both are characterized by continuously varying power laws. The predicted nonlinear behavior in \emph{time} is clearly seen in numerical simulations; these simulations also suggest nonlinear dependence on the amplitude, although we could not extract the precise form of the amplitude-dependence. A feature that is distinctive to the MBL phase is the existence of a broad parameter regime in which linear-response theory breaks down---i.e., the \emph{transient} response to driving changes its character---although the eigenstates of the Floquet Hamiltonian remain localized. This intermediate regime shrinks to a point as the MBL transition is approached (Fig.~\ref{lengths}, inset): there, the breakdown of linear response coincides with the breakdown of the Floquet-MBL steady state. (We are assuming here that $s\zeta = 1$ at the MBL transition, as conjectured in Ref.~\cite{mbmott}. It is also possible that the transition occurs for $s\zeta < 1$, in which case a small intermediate regime would persist at the transition.)

Although, for reasons of numerical tractability, we worked in the infinite-temperature limit and with one-dimensional systems, we expect that the same regimes of heating should exist throughout the MBL phase regardless of temperature or of dimensionality. We emphasize that since most of our discussion has concerned the dynamics relatively deep in the MBL phase, it is not expected to be sensitive to finite-size effects until very late times (specifically, times on the order of $\exp(L/\zeta)$ where $L$ is the linear dimension of the system). Thus, in experiments finite-\emph{time} effects, such as dissipation, are likelier to pose a challenge for our schemes than finite-size effects. Because the distinction between the Landau-Zener and Mott regimes is a generic feature of response in MBL systems, we expect that alternative time-dependent probes, such as modulation spectroscopy~\cite{tokuno2011}, will also be able to see the differences between the various regimes.

It is natural to ask about the fate of this linear-to-nonlinear response crossover \emph{beyond} the MBL transition, i.e., in the subdiffusive thermal Griffiths phase. We now briefly discuss this at a qualitative level. Suppose the undriven system is in its Griffiths phase. Then its transport is bottlenecked by rare regions that are locally ``in the MBL phase''. However, when one drives the system at large $A/\omega$, some fraction of these rare regions become delocalized by the drive (because they locally satisfy the condition that $A^{1 - s\zeta/2} \sim \omega^{s \zeta/2} W^{1 - s \zeta}$). Thus, they cease to act as bottlenecks unless their local $s\zeta$ is sufficiently small. As one continues to increase $A/\omega$, an increasing fraction of rare regions delocalize, until eventually the remaining bottlenecks become too sparse to prevent regular diffusion. Thus our results directly imply that the $A \rightarrow 0$ and $\omega \rightarrow 0$ limits fail to commute in the thermal Griffiths phase as well as the MBL phase: taking $A \rightarrow 0$ first gives anomalous diffusion whereas taking $\omega \rightarrow 0$ first gives regular diffusion. Our findings thus suggest the schematic phase diagram of Fig.~\ref{pd}, which shows  the linear and nonlinear response regimes as a function of disorder strength and the ratio of the drive amplitude and frequency. Driving a system in the MBL phase with increasingly strong fields leads to a transient crossover from linear to non-linear response (dashed lines), which need not coincide with the dynamical steady-state transitions of the Floquet Hamiltonian from a localized phase, to a subdiffusive Griffiths phase, and finally a diffusive phase (solid lines). Up to logarithmic corrections, our analysis suggests that the crossover from linear to nonlinear response should occur at $A \sim \omega$ throughout the MBL phase, including at the critical point and in the thermal Griffiths phase. This result for the critical behavior is natural~\footnote{D.A. Huse, private communication} if we take the critical point to be an infinite-randomness one, as suggested in Refs.~\cite{PalHuse, VHA, PVPtransition}: the \emph{voltage} typically has the scaling dimension of frequency~\cite{ffh}, and given infinite-randomness scaling (which suggests the characteristic length-scale for frequency $\omega$ goes as $\log \omega$), the electric field has the same scaling dimension. 

An important question for future work is how dissipation affects the dynamical regimes we have identified. In the presence of dissipation, the system is always ``thermal'' at sufficiently long times, in the sense that localization is destroyed~\cite{ngh}. In general, the system will reach a \emph{steady state}, in which the energy gained from the drive is balanced by the energy lost to the bath~\cite{rosenow_nonlinear}. Here, in addition to the drive amplitude and frequency, the dissipation rate $\gamma$ (computed, e.g., using the Golden Rule~\cite{ngh}) is crucial. When $A/\omega \ll 1, A/\gamma \ll 1$, the steady-state conductivity will coincide with the linear-response conductivity. When $A/\omega \ll 1$ but $A/\gamma \gg 1$, saturation will set in on timescales fast compared with decay; this will cause the steady-state conductivity to \emph{decrease}, and eventually to vanish as $\gamma \rightarrow 0$~\cite{rosenow_nonlinear}. An approximate estimate of the steady-state conductivity in this regime~\cite{rosenow_nonlinear} is $\sigma_{\mathrm{ss}}(\omega) \sim \sigma(\omega) (\gamma/A)$, since absorption is only possible $\gamma/A$ of the time.

One can directly extend this idea to estimate the steady-state conductivity for weakly dissipative systems in the nonlinear regime, by substituting $\gamma \sim 1/t$ in our results for the time-dependent energy spread $(\Delta E)^2(t)$, and then dividing this steady-state energy absorption by $A^2$. Thus, for strong drives or near the transition, our arguments suggest that the steady-state nonlinear conductivity is a continuously varying power law of the system-bath coupling. 
Confirming this conjecture numerically would, however, require detailed master-equation simulations~\cite{levi2015survives, *fischer2015dynamics} that are outside the scope of the present work.
Beyond these quantitative features, we expect that the steady state of the driven dissipative system will have a highly inhomogeneous temperature profile in the linear-response regime (with hot spots near resonances), but become relatively homogeneous at strong drive when the Floquet Hamiltonian is thermal. Understanding these crossovers is an important step for a full dynamical characterization of the MBL phase. 

\textit{Note added.---}As this manuscript was being prepared, we became aware of other numerical studies of the dynamical response in strongly driven many-body localized systems~\cite{Kozarzewski,Pollmann}, as well as a related, as yet unpublished, study of the response ``phase diagram'' of driven MBL systems~\footnote{J.T. Chalker, V. Khemani, and S.L. Sondhi, unpublished; see also the Nov. 2015 KITP talks by V. Khemani, http://online.kitp.ucsb.edu/online/mbl-c15/khemani/rm/jwvideo.html, and by S. Gopalakrishnan, http://online.kitp.ucsb.edu/online/mbl-c15/gopalakrishnan/rm/jwvideo.html.}.

\section{Acknowledgments}
 
We thank D. Abanin, I. Bloch, P. Bordia, B. DeMarco, M. Heyl, D. Huse, H. L\"uschen, I. Martin, R. Nandkishore, V. Oganesyan, S. Parameswaran, F. Pollmann, and U. Schneider for helpful discussions. We thank D. Huse for a critical reading of the manuscript. 
S.G. acknowledges support from the Walter Burke Institute at Caltech and from the National Science Foundation under Grant No. NSF PHY11-25915.  M.K. acknowledges support from Technical University of Munich - Institute for Advanced Study, funded by the German Excellence Initiative and the European Union FP7 under grant agreement 291763. 
E.D. acknowledges support from 
the Harvard-MIT CUA, NSF Grant No. DMR-1308435,
AFOSR Quantum Simulation MURI, 
the ARO-MURI on Atomtronics, ARO MURI Quism program,  
the Simons foundation, the Humboldt Foundation, Dr.~Max R\"ossler, the Walter Haefner Foundation, and the ETH Foundation.

\appendix

\section{Diffusion across concentration gradients} \label{gradients}

The most commonly considered case of biased diffusion is that in which particles are subjected to noise (which causes diffusion) as well as a deterministic force, such as an electric field (which causes drift). The situation considered here is somewhat different. We are concerned with the random walk of a ``particle'' (i.e., an initial configuration) in a high-dimensional configuration space. The random walk itself is unbiased, in the sense that the rates for energy-increasing and energy-decreasing transitions mediated by the drive are identical. However, the gradient comes in via the \emph{initial conditions}: lower-energy configurations are slightly more likely to be occupied at $t = 0$, when the drive is switched on. Since the driven dynamics itself is ``unbiased'' it is equally likely to heat or cool the system on any cycle; thus over time the driven system tends to ``forget'' its initial gradient. (When the Floquet Hamiltonian is thermal this causes heating to infinite temperature; when the Floquet Hamiltonian is localized, most degrees of freedom are unaffected by the drive, but the few responsive degrees of freedom precess with random phases.) 

To make this idea more concrete, we assume that heating occurs via \emph{local} processes, and that each region of the system (above a certain characteristic size $L$) heats up independently. This assumption is manifestly valid in the MBL phase; we also believe it to be valid deep in the thermal phase. We take the temperature $T$ to be greater than $L W$, where $W$ is the single-particle bandwidth. This allows us to linearize the Boltzmann factors for the various states in the system as $\exp(-E_m/T) \simeq 1 - E_m / T$. This linear energy-dependence of Boltzmann factors maps onto a linear concentration gradient in the energy-space diffusion problem. Note that the boundedness of the energy spectrum maps on to the finite extent of space over which the concentration gradient is present.

A straightforward application of these ideas is to a generic two-state system, with states labeled 1 and 2 (having energies $E_1$ and $E_2$ and occupation probabilities $P_1$ and $P_2$). The master equation for $P_1$ reads $\dot{P}_1 = \Gamma_{21} P_2 - \Gamma_{12} P_1$, where the $\Gamma$'s are intrinsic transition rates. Since these rates are unbiased (as discussed above), we have $\dot{P}_1 = \Gamma (P_2 - P_1)$, and similarly $\dot{P}_2 = \Gamma (P_1 - P_2)$. Subtracting these rates, we have that

\beq
\frac{d(P_1 - P_2)}{dt} \simeq -\Gamma(P_1 - P_2).
\eeq
so the initial concentration gradient decays at a rate $\Gamma$, which is also evidently the rate of ``energy spread'' in this two-site example, as it is the rate at which the system undergoes transitions between configurations (``sites'') of definite energy.

\section{Theory of mostly diabatic Landau-Zener crossings \label{app:LZ}}

For a given crossing one can rewrite the time-dependent Hamiltonian in a rotating frame in the form (see App. C of Ref.~\cite{shevchenko2010})

\beq\label{shev}
H' = \sum_n \Delta\sqrt{\omega/A} [\exp(- i n \omega t) \sigma_+ + \text{h.c.}] + \epsilon_0 \sigma_z
\eeq
The sum over $n$ is cut off on a scale $n \simeq A/\omega$. The matrix element $\Delta$ is the bare hopping at the scale of the particular TLS, $\Delta \sim W \exp(-n/\zeta)$. We have assumed $\omega \ll A$ as Landau-Zener transitions are important chiefly in this regime. At a large distance $x$, the first term in Eq.~\eqref{shev} can be treated perturbatively in the spirit of the rotating-wave approximation. The bandwidth of states involved in LZ transitions at this distance $\sim A x$, and there are $n(x) \sim A x/\omega$ harmonics within this window. States that lie within $\Delta(x) \sqrt{\omega/A}$ of one of these $n(x)$ harmonics of the drive frequency are resonant in the Floquet picture, and cause transport. When $x$ is relatively small, $\Delta(x) \sqrt{\omega/A} \agt A x / n(x) = \omega$. Thus, different harmonics overlap, and any transition within the drive bandwidth $A x$ occurs (as the LZ picture would predict). However, when $x$ is large and $\Delta(x)$ is correspondingly small, the inequality is flipped, and most transitions that are ``allowed'' on a naive LZ analysis are in fact off-resonant and do not contribute to transport. Thus, the MBL phase is stable against extremely long-distance LZ transitions. These transitions can instead be treated using a straightforward generalization of the Rabi-formula approach in the main text, with the matrix element $\sim A$ replaced by $\Delta \sqrt{\omega/A}$.

\end{document}